\documentclass[twoside,fleqn]{article}
\usepackage{espcrc2}

\newcommand{\AmS}{{\protect\the\textfont2
  A\kern-.1667em\lower.5ex\hbox{M}\kern-.125emS}}

\title{Topology, chiral and screening transitions at finite
       density in two colour QCD}

\author{B. All\'es\address{INFN, Sezione di Pisa, Pisa, Italy}\thanks{
        Speaker at the conference.},
        M. D'Elia\address{Dipartimento di Fisica, Universit\`a di Genova
        and INFN, Genova, Italy}
        and 
        M. P. Lombardo\address{INFN, Laboratori Nazionali di Frascati,
        Frascati, Italy}}
       
\begin{document}

\begin{abstract}
The behaviour of the topological
susceptibility in QCD with two colours and
8 flavours of quarks is studied at nonzero
temperature on the lattice across the finite
density transition. It is shown that its
signal drops at a (pseudo--)critical chemical potential $\mu_c$.
The Polyakov loop and the chiral condensate
undergo their transitions at the same value.
Pauli blocking supervenes at a value of
the chemical potential larger than $\mu_c$.
\end{abstract}

\maketitle

\section{INTRODUCTION}

\begin{figure}[htb]
\vspace{4.8cm}
\includegraphics{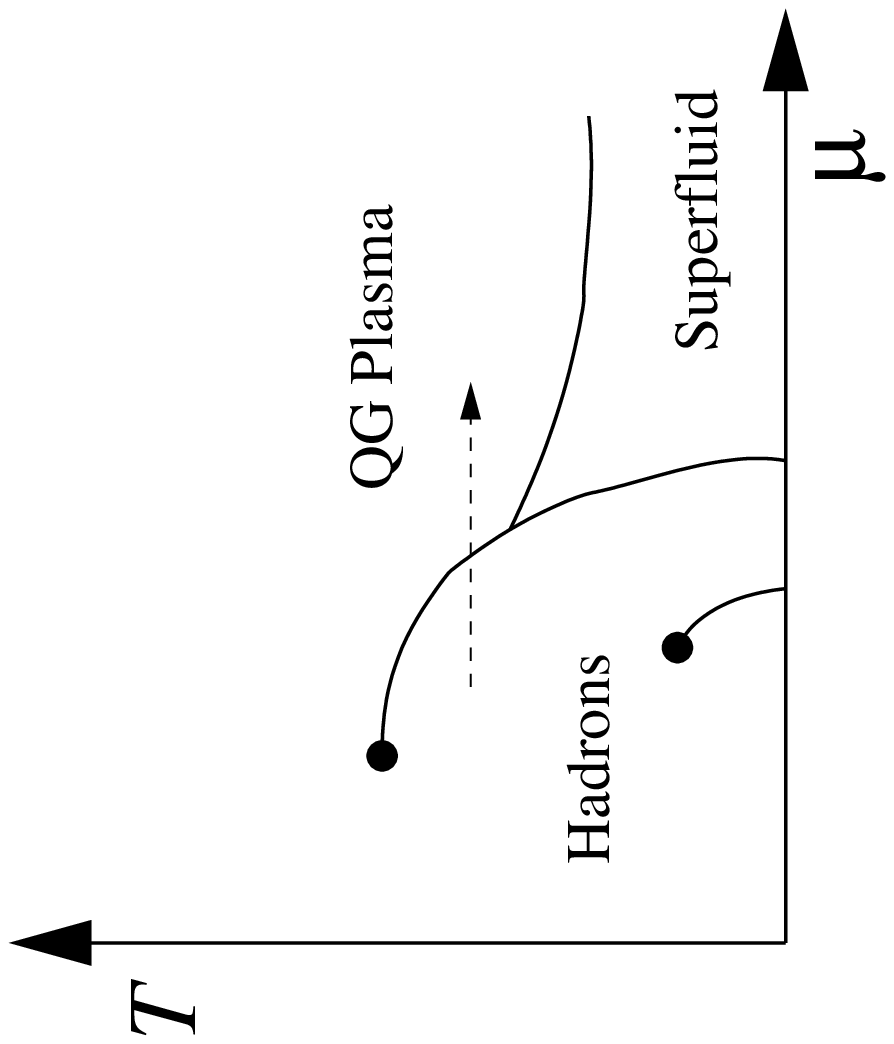}
\caption{Sketch of the $T$--$\mu$ phase diagram in QCD with two
colours. We have studied the transition crossed by the dashed
arrow at finite temperature as the density is varied.}
\label{Fig1}
\end{figure}

The phase space of QCD in the temperature and quark chemical
potential $\mu$ plane is shown schematically in Fig.~\ref{Fig1}.
Many of the transition lines and phases displayed in the Figure
are theoretical predictions with little experimental verification.
We have studied the transition as the chemical potential is
varied (dashed arrow in Fig.~\ref{Fig1}) at a fixed
temperature $T$ that is most likely above the region
where a diquark condenses.

We wanted to understand the fate of the topological susceptibility
$\chi$ across the transition and to decide whether its possible
change occurs at the same value of $\mu$ where the vector chiral
symmetry is restored and the Polyakov loop signal rises~\cite{shuryak}.

It is known that at the finite temperature transition the
signal of $\chi$ drops abruptly for Yang--Mills theory~\cite{alles1,alles2}
as well as for QCD with several values of the flavour number
$N_f$~\cite{alles3}. The present is the first work where a
similar study is performed at the finite density
transition (the full account can be found in
Ref.~\cite{heplat0602022}).

We have performed the study on the lattice by simulating the
theory with two colours in order to avoid the sign problem
that in the three colour theory makes the importance sampling
method impracticable (for methods to overcome this problem
see~\cite{filipsen} and references therein).
We expect that this modification of true QCD
has little impact on the results concerning gluon properties
and in particular instanton physics. The main differences between
two and three colour QCD regard aspects of the theory that we
have not studied, like the hadron spectrum and the
diquark condensed phase~\cite{hands,kogut}. In fact in the two
colour theory mesons are degenerate with baryons (which are made of
two quarks) and the diquark condensate phase becomes superfluid
instead of superconducting because two quarks in two colour QCD can
create a colour blind condensate.

\section{SIMULATION AND OBSERVABLES}

We have simulated the model on a $14^3\times 6$ lattice
at an inverse bare lattice gauge coupling $\beta=4/g^2=1.5$ and
quark mass $am=0.07$ in units of lattice spacing $a$.
The Wilson action was used for gauge fields and
the standard action for 8 flavours of quarks in the staggered
formulation.

The Hybrid Molecular Dynamics algorithm was chosen
to update configurations. They we-\break re separated typically
by 50--100 steps of~the algorithm in order to well
decorrelate the topology~\cite{boyd,lippert,aoki}.
We followed the behaviour of several observables
across the transition shown
in Fig.~\ref{Fig1}: chiral condensate
$\langle \overline{\psi}\psi\rangle$, Polyakov loop or Wilson
line $P$, topological susceptibility $\chi$, baryonic
density $\rho_B$ and average plaquette
${\rm Tr}\,\sqcap\hspace{-0.302cm}\sqcup /2$

\subsection{Topology}

The measurement of the topological susceptibility
requires a careful treatment of composite operators.
This quantity is defined as
\begin{equation}
\chi\equiv\int{\rm d}^4x\, \langle {\rm T}\left\{
Q(x) Q(0)\right\} \rangle\;,
\label{chi_continuum}
\end{equation}
where $Q(x)$ is the topological charge density operator
that appears in the r.h.s. of the axial singlet anomaly. $\chi$
is related to the $\eta'$ mass through the Witten--Veneziano
mechanism~\cite{witten,veneziano} and to the chiral
condensate in the massless limit~\cite{hansen}.

Eq.(\ref{chi_continuum}) can be rewritten in a more
compact form,
\begin{equation}
\chi = \frac{\langle Q^2 \rangle}{V}\;,
\label{chi_compact}
\end{equation}
where the total topological charge $Q$ is defined as
$Q\equiv\int{\rm d}^4x\, Q(x)$ and $V$ is
the spacetime volume. This expression has only
a formal meaning because it must be supplemented with
a multiplicative renormalization for each power of
the topological charge operator $Q$~\cite{campo1,alles4}
as well as a careful treatment of the contact divergences that
appear when the two operators are evaluated at the
same spacetime point~\cite{campo2}.

A detailed description of the procedure that we have
followed to extract $\chi$ can be found in~\cite{heplat0602022}.
In a nutshell, one has to define a lattice regularization
$Q_L(x)$ of the topological charge density operator.
The corresponding total lattice topological charge
is $Q_L=\sum_x Q_L(x)$ and the lattice topological
susceptibility is $\chi_L=\langle  Q_L^2 \rangle/V$.
This quantity is related to the physical susceptibility
$\chi$ by the general expression~\cite{campo1,alles4,campo2}
\begin{equation}
\chi_L = Z^2 a^4 \chi + M\;,
\label{chiL}
\end{equation}
where $Z$ is a multiplicative renormalization of $Q_L$,
$a$ is the lattice spacing and $M$ is an additive renormalization
constant. In presence of composite operators the usual
renormalization program in a renormalizable theory
is not enough to fully renormalize a Green's function. This
is the origin of $Z$ (even in the case of the pure gauge theory
where the topological charge operator has no anomalous
dimensions). On the other hand part of the contact divergences
that appear in the product of the two topological charge operators must
be subtracted. This subtraction is defined as $M\equiv\chi_L\vert_{Q=0}$,
i.e. $M$ is the value of $\chi_L$ in the sector of zero
topological charge. This condition guarantees the obvious
requirement that $\chi$ vanishes in that sector.

\subsection{Calculation of $Z$ and $M$}

One has to calculate the two constants $Z$ and $M$ and insert them
into Eq.(\ref{chiL}) in order to extract $\chi$. We shall present
the results in the form of the ratio $\chi(\mu)/\chi(\mu=0)$
as a function of $\mu$. Therefore we need not calculate $Z$ and
only $M$ is required. This additive renormalization constant was
computed by using a non--perturbative
method~\cite{vicdd,gunduc2284,alles2}. It is enough
to calculate it for one single value of the chemical potential
because $M$ is independent of infrared properties.

\begin{figure}[htb]
\vspace{4.3cm}
\includegraphics{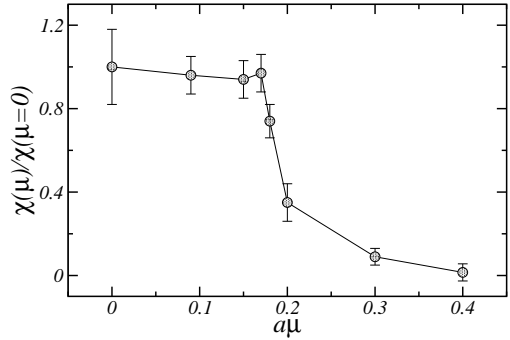}
\caption{The ratio $\chi(\mu)/\chi(\mu=0)$ as a function
of $a\mu$ showing the abrupt drop of the signal of the
topological susceptibility at a (pseudo--) critical value
of the chemical potential.}
\label{Fig2}
\end{figure}

\section{RESULTS}

In Fig.~\ref{Fig2} we show the resulting behaviour of the
topological susceptibility as a function of the chemical
potential. A clear cut drop is seen at the position
$a\mu_c=0.175(5)$. This sudden fall is an indication of
the effective restoration of the axial singlet symmetry~\cite{pw}.

\begin{figure}[htb]
\vspace{4.3cm}
\includegraphics{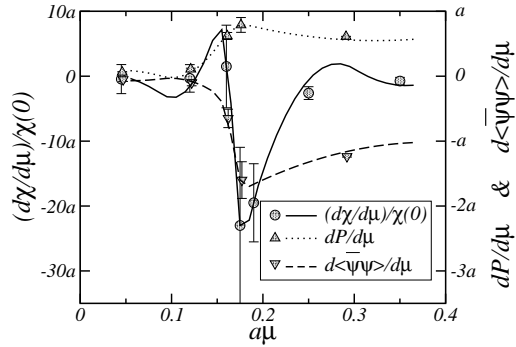}
\caption{Derivatives with respect to the chemical
potential $\mu$ of the normalized topological susceptibility
(circles, continuous line and left vertical axis), Polyakov loop
(up triangles, dotted line and right vertical axis) and chiral
condensate (down triangles, dashed line and right vertical axis).
The two vertical axes are expressed in units of
the lattice spacing $a$. The three sets of data have been slightly
shifted horizontally to avoid the overlapping of various symbols and
error bars. Lines are the result of a spline interpolation.}
\label{Fig3}
\end{figure}

We do not show the analogous figures for the Polyakov loop
and chiral condensate behaviours (see Ref.~\cite{heplat0602022}).
Instead in Fig.~\ref{Fig3} the derivatives of the three
quantities (topological susceptibility, Polyakov loop and
chiral condensate) are superimposed to give evidence of the
coincidence of the three transitions: they all happen
at the same (pseudo--)critical value $a\mu_c=0.175(5)$.
Single points are the result of the discrete derivatives
computed from the measured data while the lines
are the derivatives of the interpolations
made with a natural cubic spline on each data set.
Since the theory deconfines at the same $\mu$ where $\langle
\overline{\psi}\psi \rangle$ vanishes, the path in the $T$--$\mu$
diagram followed in our study possibly lies above the superfluid
phase~(as indicated in Fig.~\ref{Fig1})~\cite{hands22}.

\subsection{Numerical results}

In order to give an estimate of the main results in physical
units, we have done a second simulation of the model at zero
chemical potential by varying the inverse bare lattice coupling $\beta$
on the same lattice volume and have measured the Polyakov loop and
chiral condensate. The data clearly reflected the existence of a
critical $\beta_c=1.594(6)$. It was then assumed that the
corresponding transition temperature $T_c$ lies in the range 100--200
MeV. Hence a value for the lattice spacing at $\beta_c$ was
obtained. This value was then run, by making use of the two--loop beta
function, from $\beta_c$ to our working $\beta=1.5$ thus 
obtaining the lattice spacing at this beta $a(\beta=1.5)=
0.64(4)(^{+33}_{-16})$~fm (errors derived from the imprecision on $\beta_c$
and from the inaccuracy of the estimate for $T_c$ respectively).

This result allowed us to assign physical units to the
various parameters of our simulation: $\mu_c=54(2)(4)(18)$~MeV
(errors are respectively due to the estimate of $a\mu_c$,
the imprecision on $\beta_c$ and the rough approximation of $T_c$)
and $T=51(4)(17)$~MeV (error on $\beta_c$ and on $T_c$ respectively).

Notice that the numerical value of $\mu_c$ is roughly compatible
with half of the mass of the lightest baryon
(in the two colour theory it is made of two quarks
and degenerated with the pion).
However we take the above numbers with caution because
of the large systematic errors derived from the small and
coarse lattice volume,
the (wrong) gauge group and the inexact updating algorithm.

\subsection{Pauli blocking}

\begin{figure}[htb]
\vspace{4.3cm}
\includegraphics{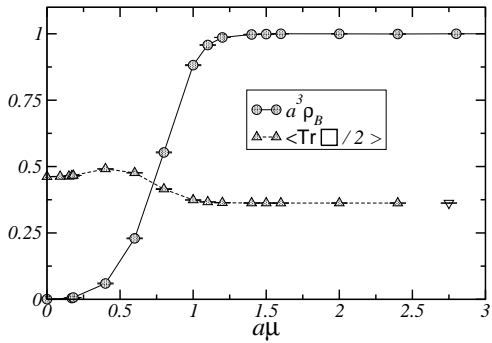}
\caption{Baryon density per space site
and average plaquette as a function of $a\mu$.
The down triangle is the result of a separate
simulation for the pure gauge theory.}
\label{Fig4}
\end{figure}

By repeating the simulation at very high values of the
chemical potential, in principle one could test the perturbative
region of the phase diagram. However the advent of the so--called
Pauli blocking makes that not viable.
As $\mu$ increases, more and more fermions
are placed in the lattice volume. Since it contains a
countable number of sites, the Pauli principle imposes a maximum allowed
number of fermions that can be placed on them. In our case
the maximum allowed density is one baryon per site~\cite{kls}.
This maximum is attained
for $a\mu_s\approx 1.2$. After that point, fermions are
frozen and the theory becomes entirely quenched. We have
verified this statement by calculating the baryon density
and the average plaquette. Their data are shown in Fig.~\ref{Fig4}.
$\rho_B$ becomes 1 and stays constant for all $\mu >\mu_s$.
Also the average plaquette stays constant after $\mu_s$ and this
constant coincides with the value obtained from a
separate Monte Carlo study where the pure gauge theory was simulated
(represented by the down triangle in Fig.~\ref{Fig4}).
We stress that Pauli blocking is a lattice
effect with no counterpart in the continuum.

\end{document}